\documentstyle[preprint,aps,floats,epsfig]{revtex}

\def\book#1[[#2]]{{\it#1\/} (#2).}
\def\am#1 #2 #3.{{\it Ann.\ Math.\ \bf#1} #2 (#3).}
\def\apj#1 #2 #3.{{\it Astrophys.\ J.\ \bf#1} #2 (#3).}
\def\atmp#1 #2 #3.{{\it Adv.\ Theor.\ Math.\ Phys.\ \bf#1} #2 (#3).}
\def\cmp#1 #2 #3.{{\it Commun.\ Math.\ Phys.\ \bf#1} #2 (#3).}
\def\comnpp#1 #2 #3.{{\it Comm.\ Nucl.\ Part.\ Phys.\  \bf#1} #2 (#3).}
\def\cqg#1 #2 #3.{{\it Class.\ Quant.\ Grav.\ \bf#1} #2 (#3).}
\def\epl#1 #2 #3.{{\it Europhys.\ Lett.\ \bf#1} #2 (#3).}
\def\grg#1 #2 #3.{{\it Gen.\ Rel.\ Grav.\ \bf#1} #2 (#3).}
\def\jmp#1 #2 #3.{{\it J.\ Math.\ Phys.\ \bf#1} #2 (#3).}
\def\ijmpd#1 #2 #3.{{\it Int.\ J.\ Mod.\ Phys.\ \bf D#1} #2 (#3).}
\def\mpla#1 #2 #3.{{\it Mod.\ Phys.\ Lett.\ \rm A\bf#1} #2 (#3).}
\def\ncim#1 #2 #3.{{\it Nuovo Cim.\ \bf#1\/} #2 (#3).}
\def\npb#1 #2 #3.{{\it Nucl.\ Phys.\ \rm B\bf#1} #2 (#3).}
\def\phrep#1 #2 #3.{{\it Phys.\ Rep.\ \bf#1\/} #2 (#3).}
\def\pla#1 #2 #3.{{\it Phys.\ Lett.\ \bf#1\/}A #2 (#3).}
\def\plb#1 #2 #3.{{\it Phys.\ Lett.\ \bf#1\/}B #2 (#3).}
\def\pr#1 #2 #3.{{\it Phys.\ Rev.\ \bf#1} #2 (#3).}
\def\prd#1 #2 #3.{{\it Phys.\ Rev.\ \rm D\bf#1} #2 (#3).}
\def\prl#1 #2 #3.{{\it Phys.\ Rev.\ Lett.\ \bf#1} #2 (#3).}
\def\prs#1 #2 #3.{{\it Proc.\ Roy.\ Soc.\ Lond.\ A.\ \bf#1} #2 (#3).}

\font\cat=cmr7
\def\E{\text{\cat E}}

\newcommand{\be}{\begin{equation}}
\newcommand{\ee}{\end{equation}}
\newcommand{\bea}{\begin{eqnarray}}
\newcommand{\eea}{\end{eqnarray}}
\newcommand{\bml}{\begin{mathletters}}
\newcommand{\eml}{\end{mathletters}}

\begin{document}
\preprint{LPT-ORSAY 02-07, hep-th/0202107}
\draft

\title{General Gauss-Bonnet brane cosmology}
\author{Christos Charmousis\footnote{E-mail address:
\texttt{Christos.Charmousis@th.u-psud.fr}} and
Jean-Fran\c{c}ois Dufaux\footnote{E-mail address: 
\texttt{Dufaux@th.u-psud.fr}}}
\address{LPT, Universit\'e de Paris-Sud, B\^at. 210, 91405 Orsay CEDEX, France}
\date{\today}
\setlength{\footnotesep}{0.5\footnotesep}
\maketitle

\begin{abstract}
We consider 5-dimensional spacetimes of constant 3-dimensional spatial
curvature in the presence of a bulk cosmological constant. 
We find the general solution of such a configuration in the 
presence of a Gauss-Bonnet term. Two classes of non-trivial 
bulk solutions are found.
The first class is valid only under a fine tuning relation 
between the Gauss-Bonnet
coupling constant and the cosmological constant of the bulk
spacetime. The second class of solutions are static and are the
extensions of the AdS-Schwarzchild black holes. Hence in the absence of
a cosmological constant or if the
fine tuning relation is not true, the generalised 
Birkhoff's staticity theorem holds even
in the presence of Gauss-Bonnet curvature terms. We examine 
the
consequences in brane world cosmology obtaining the generalised
Friedmann equations for a perfect fluid 3-brane and  discuss how this
modifies the usual scenario. 
\end{abstract}
\pacs{PACS numbers: 04.50.+h, 11.25.Mj, 11.25.Db \hfill hep-th/0202107}    

\section{Introduction}

The intriguing possibility that our Universe is only part of a higher
dimensional spacetime \cite{RSA}, \cite{EXO} 
has raised a lot of interest in the physics
community recently \cite{HW}, \cite{LOSW}, \cite{LOW} 
\cite{add}, \cite{BDL}, \cite{rs}. 
In particular 5 dimensional brane Universe models
and their cosmology have been extensively studied (see for example 
\cite{BDEL}, \cite{krauss}, \cite{prc}). 
The Universe in this
case is a gravitating homogeneous and isotropic  
brane or domain wall evolving in a constant negative curvature
spacetime. 

Amongst the interesting features of such a toy-model configuration 
is that it
verifies a generalised version of Birkhoff's staticity
theorem \cite{krauss}, \cite{prc}: a constant curvature spacetime of
constant 3-space curvature 
is locally
static; more specifically an ADS-black hole solution \cite{kottler}, 
\be
\label{kottler}
ds^2=-h(r)dt^2+h^{-1}(r)dr^2+r^2 \left({d\chi^2\over
1-\kappa \chi^2}+\chi^2d\Omega_{II}^2\right)
\ee
where $h(r)=\kappa-{\mu\over r^2}+k^2r^2$, with $\kappa=0,\pm 1$ and
$\mu$, $k^2$ are related to the black hole mass and bulk cosmological
constant respectively. To ensure the validity of this theorem it 
is essential firstly that the brane Universe is of
co-dimension 1 i.e. a domain wall type defect and secondly 
that the brane Universe is homogeneous and isotropic.  
The theorem in turn
implies a certain number of physical properties for the
configuration, in particular that the only 
dynamical
degree of freedom  
is the wall's trajectory or equivalently, for the
4-dimensional observer stuck on the brane, the expansion rate of the 
Universe. Hence although we have introduced an extra dimension,  
the number of dynamical degrees of freedom 
does not alter with respect to standard 
4 dimensional FLRW cosmology. Just like in 4 dimensional cosmology, 
once given
an equation of state relating energy density and pressure one obtains
the expansion rate or equivalently the brane Universe
trajectory. It follows rather elegantly, 
\cite{prc}, that it is totally equivalent  to
study a brane Universe evolving in a static background in the manner of
\cite{krauss} to a fixed brane
Universe in a time-dependant background \cite{BDEL}. 

When treating higher dimensional gravity theories we should keep in mind
that 4 dimensional gravity is quite special
for a numerous number of reasons. For example $D=4$ gives the minimal
number of dimensions where the graviton is non trivial and has
exactly two polarisation degrees of freedom whereas at the same time
gauge interactions of the Standard Model 
are renormalisable. 
Another special property of 4
dimensional gravity  
is the uniqueness of the Einstein-Hilbert action.

In
$D>4$, however, the situation is quite different. In 5 dimensions, in
order to
obtain the most general {\it unique} action, i.e. giving rise to
 a second order symmetric and divergence-free tensor, 
and to field equations that are of second order in the metric
components,
 {\it we have to 
add} the Gauss-Bonnet term to the usual Einstein-Hilbert plus cosmological
constant action. This is part of Lovelock's theorem \cite{lovelock}. 
Furthermore in an effective action approach of string theory, 
the Gauss-Bonnet term corresponds to the
leading order quantum correction to gravity in particular in the case of
the heterotic string \cite{gross}. The Gauss-Bonnet coupling constant is
related to the Regge slope parameter or string scale. 
Furthermore one of the important properties of string
theories is that they contain no ghosts. Interestingly  as was
demonstrated in \cite{zwiebach} the {\it only} curvature squared terms to
give ghost-free self-interactions for the graviton (around flat
spacetime) is precisely the
Gauss-Bonnet combination. 

The reason for all these nice properties shared by the Einstein-Hilbert
and the Gauss-Bonnet terms can be understood 
from a purely
geometrical 
point of view. 
The Gauss-Bonnet term is the generalised Euler characteristic of a 4
dimensional spacetime. It yields 
in $D=4$ a boundary term hence a topological and
not dynamical contribution. This is a quite general and elegant
fact. Indeed we remind the reader that in a similar fashion the
Einstein-Hilbert action in 2 dimensions 
is related to the usual Euler characteristic
of a 2-dimensional manifold.  
For example in string field theory 
the Euler characteristic $\chi$ is related to
the string coupling constant $g_s$, governing ``surface 
diagrams'' in the perturbative regime. 
In general every spacetime of even dimension $2n$ is accompanied by its
generalised Euler characteristic; which we have to
add to the gravitational action of a $2n+1$
manifold in order to
preserve uniqueness. Thus for example in 10
dimensions one has the Euler characteristics of 0 (cosmological constant), 
2 (Einstein-Hilbert), 4 (Gauss-Bonnet), 6, and 8
dimensional manifolds \cite{lovelock}. So from this discussion
it would seem natural to include 
the Gauss-Bonnet term in a 5 dimensional 
spacetime, all the more since we are interested in toy models merging
string theory with standard cosmology. 

Madore and collaborators have considered this term in order to
stabilise the 5th dimension in Kaluza-Klein theories \cite{madore}
whereas there was a
lot of effort in the 80's to obtain exact solutions in Gauss-Bonnet
theories in view to their relevance to quantum gravity corrections of
string theory (see for instance \cite{deser}, 
\cite{wiltshire0}
\cite{wiltshire}, \cite{wheeler}, \cite{myers}). 
More recently in the
context of brane universe  
it has been shown that the localised
graviton zero mode persists in the RS model in the presence 
of a Gauss-Bonnet term \cite{bruce}, \cite{neu}, 
\cite{meissner}. 
Cosmological consequences have also been studied in \cite{nouri}. 
However, only particular solutions in the bulk have been considered.  
Here we shall attack the problem in its full generality. 
We shall first of
all find and discuss 
the full bulk solutions, and then we shall investigate the brane
cosmology they induce. Not surprisingly Birkhoff's theorem will be 
in the centre of our 
analysis and its physical consequences.

In the next section we set up the basic ingredients of the problem. In
Section III we solve by brute force the field equations and find the
general solution for the bulk spacetime. 
In Section IV we discuss the relevance of the bulk solutions to
brane Universe cosmology in 5 dimensions. We conclude in section V.

\section{General setting}

Consider the following 5-dimensional action,
\be
\label{action}
S={M^3\over 2}\int d^5 x\, \sqrt{-g}\left[R+12 k^2 +\alpha 
(R_{\mu\nu\gamma\delta}R^{\mu\nu\gamma\delta}
-4R_{\mu\nu}R^{\mu\nu}+R^2)\right],
\ee
where $M$ is the fundamental mass scale of the 5-dimensional theory, 
$\Lambda=-6k^2$ is the negative bulk cosmological constant and
the Gauss-Bonnet coupling constant $\alpha$ of dimension $(length)^2$ 
,which we leave free, 
is the additional physical parameter 
of the problem. Setting $\alpha=0$ we
obviously get
the usual Einstein-Hilbert action with cosmological 
constant in 5 dimensions. 
As we discussed in the introduction, just like the 
Einstein-Hilbert action with cosmological constant is
unique in $4$ dimensions, the gravitational 
action (\ref{action}) is unique in 5 dimensions. To put it in a nutshell,
(\ref{action}) is the most general action that will yield second order
partial differential equations with respect to the metric components 
in 5 dimensions. For
this reason and for clarity we shall restrict 
ourselves to 5 dimensions. 

Let us now consider a spacetime with constant
three-dimensional spatial curvature. A general metric can 
then be written,
\be
\label{metric}
ds^2=e^{2\nu(t,z)}B(t,z)^{-2/3}(-dt^2+dz^2)+B(t,z)^{2/3}
\left({d\chi^2\over
1-\kappa \chi^2}+\chi^2d\Omega_{II}^2\right)
\ee
where $B(t,z)$ and $\nu(t,z)$ 
are the unknown component fields of the metric and $\kappa=0, 
\pm 1$ is
the normalised curvature of the 3-dimensional homogeneous and 
isotropic surfaces. 
We choose to use the conformal gauge in order to take full advantage 
of the
2-dimensional conformal transformations in the $t-z$ plane. This is the
setup for a cosmological wall or brane-Universe of co-dimension 1. We note on
passing that a co-dimension 2 or higher set-up would have lost 
the two dimensional
conformal freedom. We will see in what follows that this freedom is
essential for the integrability of the system wether or not we include
the Gauss-Bonnet combination.

The field
equations we are seeking to solve are found by varying the above action
(\ref{action}) with respect to the background metric and read 
\bea
\label{field}
\E_{\mu\nu}&=&G_{\mu\nu}-6k^2g_{\mu\nu}-
\alpha\left[{g_{\mu\nu}\over 2}
(R_{\alpha\beta\gamma\delta}R^{\alpha\beta\gamma\delta}
-4R_{\gamma\delta}R^{\gamma\delta}+R^2)\right.\nonumber\\&-&\left.2R
R_{\mu\nu}+4R_{\mu\gamma}R^{\gamma}_{\mbox{ }\nu} 
+4R_{\gamma\delta}R_{\mbox{ }\mu\mbox{ }\nu}^{\gamma\mbox{ }\delta}
-2R_{\mu\gamma\delta\lambda}
R_{\nu}^{\mbox{ }\gamma\delta\lambda}\right]=0
\eea
where now the symmetric tensor 
$\E_{\mu\nu}$ has replaced the usual Einstein tensor 
$G_{\mu\nu}$ and is also divergence free. Taking the trace 
of (\ref{field})
one can show that for a solution, the Ricci scalar is
a multiple of the Lagrangian in (\ref{action}) (see for instance 
\cite{wheeler}). Thus
the behaviour (in particular singularities) 
of the scalar curvature $R$ is shared by the Gauss-Bonnet scalar
in (\ref{action}). Hence not surprisingly we can deduce that 
although the field equations change radically, 
spacetime curvature still plays the same
physical role for the critical points of the action (\ref{action}). 

\section{The general solution for the bulk spacetime}

Before plunging in the field equations{\footnote{For the full field
equations see Appendix.}} it is rather useful to review the
generalisation of Birkhoff's theorem in the presence of a cosmological
constant as it appeared recently in \cite{prc}. Furthermore we shall 
use
exactly the same method to derive the general solution. 

In this subcase the field
equations are obtained setting $\alpha=0$ in (\ref{action}) and read
$$
R_{\mu\nu}=-{2\Lambda\over 3}g_{\mu\nu}.
$$
There are two key ingredients in this
method. First of all in order to make use 
of the $t-z$ conformal symmetries of 
(\ref{metric})
it is important to pass to light-cone coordinates, 
\be
\label{lightcone}
u={{t-z}\over 2},\qquad v={{t+z}\over 2}.
\ee
Secondly taking the
combination $R_{tt}+R_{zz}\pm 2R_{tz}=0$, one obtains the integrability
conditions which read\footnote{From now on $B_{,u}$ 
represents the partial derivative of $B$ with respect to $u$ etc.}
\bea
\label{prc}
B_{,uu}-2\nu_{,u} B_{,u}&=&0,\\
B_{,vv}-2\nu_{,v} B_{,v}&=&0.
\eea
Note then that these
are ordinary differential equations with respect to $u$ and $v$
respectively and are independent of the physical parameter of the
problem, namely, the cosmological constant $\Lambda$. 
As their name indicates they are directly integrable giving
\be
\label{peter}
B=B(U+V)\qquad e^{2\nu}={B'}{U'}{V'}
\ee
where $U=U(u)$ and $V=V(v)$ are arbitrary functions of $u$ and $v$, 
and a prime stands for the total derivative of the function with 
respect to its unique variable.
 Using a conformal transformation,
$$
U={{\tilde{z}-\tilde{t}}\over 2},\qquad  V={{\tilde{z}+\tilde{t}}\over 2}$$
gives that the solution is locally static $B=B(\tilde{z})$ and
Birkhoff's theorem is therefore true. Starting from a general time and
space dependant metric, spacetime has been shown to be locally static or
equivelantly that there exists a timelike Killing vector field (here
$\partial\over \partial\tilde{t}$). Note that we did not have 
to find
the precise form of the solution for $B$. The integrability conditions
actually suffice
to prove staticity and thus Birkhoff's theorem. 
By use of the
remaining field equations we can then find the form 
of $B$, leading after
coordinate transformation to the
topological black hole solution (\ref{kottler}). 
Note that the solution becomes 
$\tilde{t}$-dependent as we cross the event horizon of the black
hole. For more details the reader can consult \cite{prc}.

Let us now turn to our case of interest with $\alpha\neq 0$. 
In the presence
of the Gauss-Bonnet term we can expect that if the system is indeed
integrable then some integrability equation should be
reproduced. Putting away technicalities this
is the essence of what we shall do here. 
In analogy to the previous case 
let us take the combination, $\E_{tt}+\E_{zz}
\pm 2\E_{tz}=0$. On passing to light cone coordinates
(\ref{lightcone}) we get after some manipulations
\bea
\label{integrability}
\left(9B^{4/3}e^{2\nu}+36\alpha \kappa B^{2/3} e^{2\nu}
+4\alpha B_{,u}B_{,v}\right)(B_{,uu}-2\nu_{,u} B_{,u})&=&0\nonumber\\
\left(9B^{4/3}e^{2\nu}+36\alpha \kappa B^{2/3} e^{2\nu}+
4\alpha B_{,u}B_{,v}\right)(B_{,vv}-2\nu_{,v} B_{,v})&=&0
\eea
Note how the Gauss-Bonnet terms factorise nicely leaving the
integrability equations (\ref{prc}) we had in the absence of $\alpha$. 

Let us neglect for the moment 
the degenerate case where either 
$B_{,u}=0$ or $B_{,v}=0$  
corresponding to flat solutions \cite{taub} (see Appendix). 
For $B_{,u}\neq 0$ 
and $B_{,v}\neq 0$ the situation is clear: 
either we have static solutions and Birkhoff's
theorem holds as in the case above or we will have 
\be
\label{class1}
e^{2\nu}={4\alpha(B_{,z}^2-B_{,t}^2)\over {9B^{2/3}(B^{2/3}
+4\alpha\kappa)}}
\ee

Let us first examine the latter case, that we will call Class I 
solution. 
The two remaining field equations $\E_{\chi\chi}=0$ and 
$\E_{tt}-\E_{zz}=0$ give after some algebra the simple
relation,
\be
\label{fine}
8\alpha k^2=1
\ee
This is quite remarkable: if the coupling constants obey this simple
relation (\ref{fine}) 
then the $B$ field is an {\it arbitrary} function of space and
time. Note in passing that Class I solutions exist in 
arbitrary dimension $d$ if the fine tuning relation 
${96\alpha k^2\over {(d-1)(d-2)}}=1$ is satisfied. We can already 
deduce that Birkhoff's theorem does not hold for
non zero cosmological constant. {\footnote{Note however that for a
non-zero charge $Q$ and spherical symmetry ($\kappa=1$) Birkhoff's theorem is
always true as was shown by Wiltshire \cite{wiltshire0} 
(see also \cite{wiltshire})}} 
However in the absence of a cosmological
constant it is always trivially true since (\ref{fine}) is impossible. 
Also we can note from (\ref{fine}) that 
a positive Gauss-Bonnet constant $\alpha>0$, as in heterotic string
theory,  
 demands a negative cosmological constant and vice-versa. The
Class I metric reads,
\be
\label{CL1}
ds^2={4\alpha(B_{,z}^2-B_{,t}^2)\over
{9B^{4/3}(B^{2/3}+4\alpha \kappa)}}(-dt^2+dz^2)
+B^{2/3}\left({d\chi^2\over
1-\kappa \chi^2}+\chi^2d\Omega_{II}^2\right)
\ee
under the constraint (\ref{fine}) where we emphasize that $B(t,z)$ is an
arbitrary function of $t$ and $z$. To simplify somewhat set $B=R^3$ to get,
\be
\label{CL12}
ds^2={R_{,z}^2-R_{,t}^2\over
{\kappa+{R^2\over {4\alpha}}}}(-dt^2+dz^2)
+R^2\left({d\chi^2\over
1-\kappa \chi^2}+\chi^2d\Omega_{II}^2\right)
\ee
This solution has generically a curvature singularity for
$R_{,z}=\pm R_{,t}$. The parameter $\alpha$ is related here to the 
5-dimensional cosmological constant via (\ref{fine}).
The Class I static solutions are given by,
\be
\label{CL1static}
ds^2=-{A(R)^2\over
{\kappa+{R^2\over {4\alpha}}}}dt^2+{dR^2\over\kappa+{R^2\over {4\alpha}}}
+R^2\left({d\chi^2\over
1-\kappa \chi^2}+\chi^2d\Omega_{II}^2\right)
\ee
with $A=A(R)$ now an arbitrary function of $R$. 
Time-dependent solutions for $\alpha>0$ are only possible for
$R^2< 4\alpha$ and $\kappa=-1$.{\footnote{For $\alpha<0$ the
situation is interchanged with static solutions possible only for
$\kappa=1$ and $R^2<-4\alpha$.}}

In order to obtain $t$ and $z$ dependent solutions it suffices to 
take
the functional $R$ to be a non-harmonic function. 
Take for instance $R=exp(f(t)+g(z))$,
with $f$ and $g$ arbitrary functions of a timelike and spacelike
coordinate respectively. Let us also assume $\kappa=0$ 
for simplicity,
the Class I metric in proper time reads,
\be
\label{CLtz}
ds^2=-d\tau^2+
{4\alpha dg^2\over {1+4\alpha f_{,\tau}^2}}+e^{2(f+g)}\left({d\chi^2\over
1-\kappa \chi^2}+\chi^2d\Omega_{II}^2\right).
\ee
Note here again that $f$ is an arbitrary function of time. 

On the
other hand if (\ref{fine}) does not hold then Birkhoff's theorem 
remains
true in the presence of the Gauss-Bonnet terms i.e. {\it the general 
solution assuming the presence of a cosmological constant in the bulk
and 3 dimensional constant curvature surfaces is static if and only
if (\ref{fine}) is not satisfied}. 
In this case the remaining two equations give the same ordinary
differential equation for $B(U+V)$ which after one integration reads,
\be
\label{B}
{B'}+9B^{2/3}(k^2 B^{2/3}+\kappa)+
18\alpha\left({B'\over {9B^{2/3}}}+\kappa\right)^2=9\mu
\ee
where $\mu$ is an arbitrary integration constant. Then by making $B$ 
the
spatial coordinate and setting $B^{1/3}=r$ we get the solution 
discovered and discussed in
detail by Boulware-Deser \cite{deser} 
($\kappa=1$) and Cai \cite{cai} ($\kappa=0,-1$),\footnote{We have kept
the same label as in
(\ref{metric}) for the rescaled time coordinate .}
\be
\label{BH}
ds^2=-V(r) dt^2+{dr^2\over V(r)}+r^2\left({d\chi^2\over
1-\kappa \chi^2}+\chi^2d\Omega_{II}^2\right)
\ee
where $V(r)=\kappa+{r^2\over
{4\alpha}}[1\pm\sqrt{1-8\alpha k^2+8{\alpha \mu\over r^4}}]$, and
$\mu$ plays the role of the gravitational mass. 
The maximally symmetric solutions are obtained by setting $\mu=0$.
There are two AdS branches permitted by the solution
 for $\alpha>0$ (\cite{cai}, \cite{deser}). 
We can have both 
de-Sitter and Anti-de-Sitter for $\alpha<0$. 
Generically, as shown
by Boulware and Deser \cite{deser}, 
only one of the branches is physical, the $'+'$
branch being classically unstable to small perturbations 
and yielding a graviton ghost. 
For $\alpha>0$ and the $'-'$ branch 
 there is a black hole
singularity at $r=0$, a unique event horizon and asymptotically one
approaches the 5-d topological black hole solutions 
\cite{kottler}, \cite{birmingham} (see also \cite{roberto}). 
For a general and thorough
analysis of the Gauss Bonnet black hole solutions, 
and their thermodynamics 
we refer the reader to
\cite{cai}, \cite{deser} {\footnote{Perturbative AdS black holes with
$R^2$ curvature terms 
and their thermodynamics have been discussed in \cite{nojiri})}. 
It is interesting to point out that {\it only} the planar
black hole $\kappa=0$ obeys the entropy-area formula. Indeed it turns out,
\cite{cai} that the planar black hole shares exactly the same
thermodynamic{\footnote{The interested reader could consult the more
general approach of \cite{wald}}} 
properties as the planar topological black hole
($\alpha=0$) although the two solutions differ considerably. We will
come back to this point in the next section. 
Furthermore for small $\alpha$ we have,
$V(r)=\kappa+k^2r^2(1+2\alpha k^2)
-{\mu\over r^2}[1+2\alpha (2k^2-{\mu\over r^4})]+O(\alpha^2)$
and indeed for $\alpha=0$ we get the usual Kottler solution
\cite{kottler}.

Now
notice how (\ref{fine}) is a particular 'end' point for (\ref{BH})
since the maximally symmetric solution is defined only for $1\geq 8\alpha
k^2$ (for $\alpha<0$ there is no such restriction). We can deduce in all
generality that
for $1\geq 8\alpha k^2$ there is a unique static solution (\ref{BH}).
When (\ref{fine}) is satisfied 
and $\mu=0$, 
the two branches coincide and 
$V=\kappa+{r^2\over {4\alpha}}$, 
which is then a particular  Class I solution (\ref{CL1}) for the  
value $A(r)=V(r)$. For $1\leq 8\alpha k^2$ no solutions exist.

\section{Brane world cosmology}

Having evaluated the general solution in the bulk 
we now consider a
4-dimensional 3-brane where matter is 
confined. We
furthermore suppose following the symmetries of our metric
(\ref{metric}) that matter on 
the brane is modelled by a perfect fluid of energy density $\rho$ and
pressure $p$. The brane is fixed at $z=0$, 
and the energy-momentum tensor associated with the brane takes the form,
$$
T_{\mu}^{\nu (b)}
={\delta(z)\over \sqrt{g_{zz}}}diag(-\rho(t),p(t),p(t),p(t),0).
$$
and the
field equations read,
\be
\label{branefe}
\E_{\mu\nu}=M^{-3}T_{\mu\nu}^{(b)} \delta(z).
\ee
We will assume $Z_2$ symmetry across the location of the brane at $z=0$, 
and set $M^{3}=1$ for the time being. 

Now before proceeding
there are three important points to take into account. First of all the
Israel junction conditions are no longer valid since we have included
the Gauss-Bonnet term in the gravitational action. Although the
Gauss-Codazzi integrability conditions are universal for any
 spacelike
or timelike
hypersurface (see for instance \cite{wall}), the Israel junction
conditions have to be generalised in order to take into account the
addition of the Gauss-Bonnet term \cite{nathalie} in the 
gravitational
action (\ref{action}). So in order to evaluate the brane
junction conditions we choose to bifurcate the Israel junction
conditions, integrating the field equations (\ref{branefe}) on an
infinitely small interval across the brane location at $z=0$. 

The second important point is that since the field equations are of
second order we will {\it encounter at most} 
second derivatives of $z$ and
therefore the metric component fields {\it have to be continuous}. 
Indeed first order derivatives contain a jump in the metric
given by means of the Heaviside distribution whereas second order
derivatives contain a Dirac distribution at $z=0$, to be matched with the
brane energy-momentum tensor (\ref{branefe}).
Note that had we considered any other combination of quadratic curvature
terms in the action, the situation would have been different. 
The good behavior of the Gauss-Bonnet combination 
is coherent with the fact that (\ref{action})
is unique in 5 dimensions, just as ordinary Einstein Hilbert gravity
plus cosmological constant is unique in 4 dimensions. 
Hence we can expect a regular
gravity theory and hence regular boundary conditions.

The final remark turns out to be crucial
 for the correct evaluation of
the junction conditions and has been a source of confusion in the
related literature. 
Indeed note
 that, in (\ref{branefe}), first order
derivatives with respect to $z$, multiplying second order derivatives of
the metric functions, are always encountered as squares. Thus, the first
order part
involving Heaviside distributions turns out to be equal
to $+1$ everywhere {\it except at $z=0$}. Although this is a removable 
discontinuity it occurs
in the location of the Dirac distribution and hence 
the junction conditions {\it
are not} obtained by simply matching the Dirac 
distributions in the field equations (\ref{branefe}). It is imperative 
that we integrate over an infinitely small interval across the
brane location \footnote{We thank Stephen Davis for discussions 
on this point} and then take the limit. 
In doing so 
the $t-t$ and $\chi-\chi$ components of
(\ref{branefe}) give respectively the energy density and pressure on the
brane:
\be
\label{energy}
\rho(t)=-\frac{2}{9} e^{-3\nu} B^{-2} B_{,z} (I_1 + \frac{8\alpha}{3}
B_{,z}^2)
\ee
\be
\label{pressure}  
p(t)=\frac{2}{9} e^{-3\nu} B^{-2} \left[ I_1 (B \nu_{,z}-\frac{1}{3}
B_{,z})+ B_{,z} [6 e^{2\nu} B^{4/3} + 8 \alpha B (B_{,tt}-\nu_{,t}
B_{,t}) - \frac{8\alpha}{3} B_{,t}^2] \right]
\ee
with $I_1=9B^{4/3}e^{2\nu}+36\alpha \kappa B^{2/3} e^{2\nu}
-4\alpha(B_{,z}^2-B_{,t}^2)$ (see appendix). All the functions in the
RHS of (\ref{energy}) and (\ref{pressure}) are evaluated at $z=0^+$
since we have assumed $Z_2$-symmetry. The domain wall case ($\rho=-p$)
corresponding to a Poincar\'e invariant brane has been treated in
\cite{neu}, \cite{meissner}.

We will first focus on the Class II solution in the bulk. 
In this case the junction conditions 
remain
invariant under the conformal boost $u\rightarrow f(u)$,
$v\rightarrow f(v)$ just as
for $\alpha=0$. Therefore for a fixed brane (or boundary) at $z=0$ 
there is a single degree of freedom $U'$
{\it or} $V'$ for the bulk spacetime which is 
evolving 
in time. By virtue of Birkhoff's theorem 
this is equivelant to taking a moving brane (or boundary) in the static black 
hole background.
We make use of this fact now
to pass on to the static bulk configuration (for a detailed discussion
see 
(\ref{BH})). 

Consider a brane Universe observer. The expansion parameter (or
wall's trajectory) reads,
$R(\tau)=B^{1/3}(t,0)$ whereas proper time $\tau$ is given by 
$d\tau=e^{\nu(t,0)}B^{-1/3}(t,0) dt$. For the solution (\ref{BH})
we have relations (\ref{peter}) and for example the Hubble expansion rate is
given by,
$$
H=\frac{1}{R} \frac{dR}{d\tau}=
{(U'+V'){B'}\over {6e^{\nu}B^{2/3}}}
$$
First, using (\ref{peter}), (\ref{B}), (\ref{energy}) and
(\ref{pressure}), 
we may obtain the standard 
conservation equation on the brane:
\be
\label{conserv}
\frac{d\rho}{d\tau}+ 3 H (p+\rho) =0
\ee
which is a consequence of the Bianchi and Bach-Lanczos identities 
for the Einstein tensor and Gauss-Bonnet terms respectively in 
(\ref{field}).
Then, using (\ref{peter}), (\ref{B}) and (\ref{energy}) we get the generalised 
Friedmann equation:
\bea
\label{fried}
\left( \frac{\rho}{16\alpha} \right) ^2 = \left( H^2+\frac{V(R)}{R^2}
\right) ^3  \mp C \left(
H^2+\frac{V(R)}{R^2} \right) ^2 + \frac{1}{4} C^2 \left( 
H^2+\frac{V(R)}{R^2} \right)
\eea
where we have defined
$$
C = \frac{3}{4\alpha} (1-8\alpha k^2 + \frac{8\alpha\mu}{R^4})^{1/2}
$$
and $V(R)$ is the black hole potential (\ref{BH}). This equation relates
the brane trajectory $R(\tau)$ with the energy density of the
brane. Equations (\ref{fried}) and (\ref{conserv})  
fix the unique degree of freedom in the bulk $R=R(\tau)$. In 
Einstein-Hilbert brane cosmology where $\alpha=0$,  
$\rho^2$ depends linearly on $H^2+\frac{h(R)}{R^2}$, where $h(R)$ is  
the Kottler potential (\ref{kottler}).
Indeed taking heuristicaly the limit of small Gauss-Bonnet
coupling ($\alpha \rightarrow 0$) in (\ref{fried}) yields
the usual Friedmann
equation for a 3-brane embeded into a (Einstein-Hilbert)
five-dimensional bulk (see for instance \cite{krauss}):
\be
\label{Oalpha}
H^2={\rho^2\over 36}-{\kappa\over R^2}-k^2+{\mu\over
R^4} + O(\alpha)
\ee
 for the lower $'+'$ sign in (\ref{fried}). 
This sign corresponds to the stable branch for
the black hole solution (\ref{BH}) 
as demonstrated by Boulware and Deser \cite{deser}. 
{\footnote{Note that for $\alpha \rightarrow 0$ 
the upper $'+'$ branch of (\ref{BH}) 
yields a singular negative term in
the RHS of (\ref{fried}), which ties in nicely with
the results of \cite{deser} showing the instability of this branch.}}  
At early times, (\ref{Oalpha}) leads to 
unconventional cosmology $H^2 \propto \rho^2$ \cite{BDL}, 
in contrast with 
standard four-dimensional cosmology where $H^2 \propto \rho$. Now, note
from (\ref{fried}) how Gauss-Bonnet gravity 
yields $\rho^2$ depending also on
$(H^2+\frac{V(R)}{R^2})^2$ and $(H^2+\frac{V(R)}{R^2})^3$. In general
the three powers in the RHS of (\ref{fried}) are expected to dominate
successively during the cosmological evolution of the universe. This may
have interesting consequences for early as for late time brane cosmology.
 Generically, the cosmological evolution
resulting from (\ref{fried}) strongly
depends on the epoch under consideration and on the order of magnitude
of the bulk lagragian parameters $\alpha$ and $k$. A detailed study lies
beyond the scope of a simple application of Birkhoff's theorem to higher
dimensional theories and will be undertaken in future work.

As an illustrative example we 
consider here the late time cosmology of a spatially flat ($\kappa = 0$)
expanding universe, for the particular relation between the bulk
parameters 
\be
\label{snif}
8\alpha k^2 = -3 \mbox{  ,}
\ee
We will consider the $'-'$ (upper) sign in (\ref{fried}) and expand up
to $O(\frac{1}{R^4})$ (large scale factor). In this particular case,
both the effective cosmological constant on the brane and the black hole
term 
{\footnote{This term is usually referred to as
the 'dark radiation term' due its 'radiation' like behavior. This is
however misleading since the bulk solution does not radiate 
(Birkhoff's theorem) and there
are furthermore no 'radiation' like particles in the bulk.}} in
$\frac{\mu}{R^4}$ vanish. Equation (\ref{fried}) then reduces to
$$
\frac{4}{3}\alpha H^6 + H^4 - \frac{\mu}{2 R^4} H^2 = \frac{\rho^2}{192
\alpha} + O(\frac{1}{R^8})
$$
At sufficiently late times, the leading contribution for the brane
energy density is $H^2 \propto \rho$, as in
standard four-dimensional cosmology, without the need to introduce any
brane tension.
For $\alpha > 0$ we have a positive cosmological constant in the
five-dimensional bulk action, but however the bulk space-time is AdS, 
as may be seen through the expression of the black-hole
potential $V(R)$ (\ref{BH}). 
Domain wall solutions
($\rho=-p$) without
brane tension and in the presence of higher curvature terms, 
have been studied in \cite{meissner}. It has been proved
that they may allow for a massless normalisable four-dimensional
graviton. We can regard (\ref{snif}) as a special relation involving the 
bulk parameters in contrast to the Randall-Sundrum relation relating
brane and bulk parameters.
The catch however is that we used the $'+'$ (upper) branch of
the solution in (\ref{BH}), which is unstable according to \cite{deser}. 
It would be
interesting to study the cosmology of similar cases in higher dimensions
with higher order Euler densities, which may be stable
\cite{meissner}.      

Let us now consider the case $8\alpha k^2=1$ and $\mu=0$. Then,
(\ref{fried}) becomes:
$$
H^2= (\frac{k^2}{2} \rho)^{2/3} - 2 k^2 - \frac{\kappa}{R^2}
$$
This is the result we obtain for a  Class I solution in
the bulk, by directly using (\ref{class1}) and (\ref{CLtz}) (extended
for $\kappa \neq 0$) in (\ref{energy}). In this case, with
(\ref{pressure}), one also obtains (\ref{conserv}). Again for Class I
solutions, there is only one dynamical degree of freedom in brane cosmology,
namely the function $f(\tau)$ appearing in (\ref{CLtz}).

Now it is interesting to study the late time cosmology of an expanding
brane-universe resulting from the generalised Friedmann equation
(\ref{fried}). In doing so, we take into account the tension
(vacuum energy) $T$ of the brane ($\rho \rightarrow T + \rho$), and keep
only linear terms in $\rho$ and $\frac{\mu}{R^4}$ (large scale
factor). We concentrate on the physical Class II solutions ('-' branch
in (\ref{BH})), with a negative cosmological constant in the bulk action
and with $\alpha > 0$ (as required by string theory), so that 
$0 \leq 8 \alpha k^2 << 1$.
For a zero effective cosmological constant on the brane, one
has to impose the modified Randall-Sundrum fine-tuning condition:
\be
\label{tension}
T = \left[ \frac{1-\sqrt{1-8\alpha k^2}}{\alpha} \right] ^{\frac{1}{2}} 
(2+\sqrt{1-8\alpha k^2})
\ee
which indeed allows for a Kaluza-Klein zero-mode localized on the brane
and a finite volume element, $m_{Pl}<\infty$ \cite{meissner}.
Then, up to $O(\rho)$ and $O(\frac{\mu}{R^4})$, (\ref{fried}) gives
\be
\label{friedlate}
H^2 = \frac{m_{Pl}^{-2}}{3} \rho - \frac{\kappa}{R^2} +
(2 + \sqrt{1-8 \alpha k^2})^{-1} \frac{\mu}{R^4}
\ee
where the reduced four-dimensional Planck mass is given by
\be
\label{mplanck}
m_{Pl}^{-2} = \frac{M^{-3}}{2 \sqrt{\alpha}} \left(
\frac{(1- \sqrt{1-8 \alpha k^2})^{1/2}}{2- \sqrt{1-8 \alpha k^2}} \right)
\ee
and we have restored the fundamental mass scale $M$ 
of the 5-dimensional theory. Note that the ``cosmological''
Planck mass, as defined above, agrees with the four-dimensional Newton
constant obtained 
through estimation of the static gravitational potential at long
distances along the brane \cite{neu}.

Repeating the same procedure as above in the special case $8 \alpha k^2
= 1$, one finds that (\ref{tension}) and (\ref{mplanck}) still hold,
while the $\mu$-dependent term in the Friedmann equation
(\ref{friedlate}) turns out to be in $\frac{\mu^{2/3}}{R^6}$. Thus  
for $0 <  8\alpha k^2 \leq 1$, one sees from 
(\ref{mplanck}) that:
$$
M^3 < m_{Pl}^2 k
$$
whereas strict equality holds in the absence of the Gauss-Bonnet
term (which is the leading quantum gravity correction term). 
Hence for fixed
4-dimensional Planck mass and cosmological constant in the bulk, 
``quantum corrections'' for gravity in the bulk  tend to decrease the
fundamental mass scale $M$ of the 5-dimensional theory.

\section{Conclusions}

In this paper we have studied Gauss-Bonnet brane cosmology in a 5
dimensional spacetime. 
Our main motivation for including the Gauss-Bonnet term is that 
the usual 5 dimensional 
gravitational action (\ref{action}) is then unique \cite{lovelock} 
as we noted in the Introduction. Furthermore the Gauss-Bonnet
coupling constant $\alpha$ provides a window to 
the leading quantum gravity
correction coming from string theory. Throughout our analysis the
technical difficulties induced by the inclusion of the higher order
curvature term where seen to be overcome quite elegantly for the
Gauss-Bonnet combination.

Indeed starting 
from a homogeneous and isotropic 3-space in constant bulk curvature we
found the general spacetime solutions to the field equations. 
Under a particular relation between the bulk cosmological constant and
the Gauss-Bonnet coupling,  a space and time dependant
solution (Class I) of the field equations was found. 
If however this special relation is not 
satisfied then the unique solution is the 
black hole solution discovered and 
discussed in \cite{deser}, \cite{wheeler}, \cite{wiltshire0} 
and \cite{cai}. 
Therefore quite elegantly Birkhoff's staticity theorem holds 
and all its interesting properties go through just like in the ordinary 
$\alpha=0$ case \cite{prc}. 

As a concrete application to the generalised Birkhoff's theorem
we studied brane cosmology in 5 dimensions for a 4-dimensional
perfect-fluid brane. 
As it turns out special care has to be taken
when deriving the generalised Friedmann equations for the
brane. Although there are no ill
defined distributional products (unlike any other higher order curvature
theory) a limiting procedure has to be undertaken in order to obtain the
junction conditions. On doing so it is found that the 
generalised Friedmann equation
involves a third order polynomial in $H^2$ which yields drastic changes
to conventional ($\alpha=0$) brane cosmology. Also generically  
Gauss-Bonnet gravity  tends to 
decrease the 5 dimensional fundamental mass scale, 
which is interesting if we 
interpret the Gauss Bonnet term as the 
leading string quantum gravity correction. It is now important to
investigate whether the Gauss-Bonnet term can give ordinary late time
FLRW cosmology without the usual fine tuning conditions needed in
Einstein-Hilbert brane cosmology. Work in this direction is under way.

\section*{Acknowledgements}

It is a great pleasure to thank Brandon Carter, 
Stephen Davis, Emilian Dudas and  John Madore for encouraging and 
discouraging remarks in the early stages of
this work. We thank in particular Pierre Binetruy and 
Jihad Mourad for fruitful discussions. CC also thanks
Ruth Gregory for pointing out \cite{cai} and Simon Ross.  

\noindent
{\bf Note added:}  \cite{port} has been brought 
to our attention, which appeared quasi-simultaneously to our paper 
and discusses some related issues.

\appendix\section*{Field equations}

The field equations obtained from (\ref{field}) read,
\bea
\label{appendix1}
\E_{\chi\chi}&=&-{1\over {9e^{4\nu}B}}(B_{,tt}-B_{,zz})
[12e^{2\nu}B^{2/3}(2\alpha\kappa+B^{2/3})-I_1]\nonumber\\
&+&{8\alpha\over {9e^{4\nu}}}[B_{,zz}B_{,tt}+(\nu_{,t}^2-\nu_{,z}^2)
(B_{,t}^2-B_{,z}^2)+B_{,tz}(2\nu_{,t}B_{,z}+2B_{,t}\nu_{,z}-B_{,tz})
\nonumber\\
&-&(B_{,tt}+B_{,zz})(\nu_{,t}B_{,t}+\nu_{,z}B_{,z})]\nonumber\\
&-&{20\alpha\over {81e^{4\nu}B^2}}(B_{,t}^2-B_{,z}^2)^2
-{I_1\over 9e^{4\nu}}
(\nu_{,tt}-\nu_{,zz})\nonumber\\
&-&6k^2 B^{2/3}-{\kappa\over {3e^{2\nu}B^{4/3}}}
[I_1-6e^{2\nu}B^{2/3}(6\alpha\kappa+B^{2/3})]=0
\eea
\bea
\label{appendix2}
\E_{tt}-\E_{zz}&=&{I_1\over
{9e^{2\nu}B^{7/3}}}(B_{,tt}-B_{,zz})+{12e^{2\nu}k^2\over
{B^{2/3}}}\nonumber\\
&+&{6\kappa e^{2\nu}\over B^{4/3}}-{8\alpha\over {27B^{10/3}e^{2\nu}}}
(B_{,t}^2-B_{,z}^2)^2\nonumber\\
&-&{8\alpha\kappa\over {3 B^{8/3}}}
(B_{,t}^2-B_{,z}^2)=0
\eea
The integrability conditions $\E_{tt}+\E_{zz}\pm 2\E_{tz}$ are:
\bea
\label{appendix3}
I_1(B_{,tt}+B_{,zz}+2B_{,tz}-2\nu_{,t}B_{,t}-2\nu_{,z}B_{,z}+
2\nu_{,t}B_{,z}+2B_{,t}\nu_{,z})&=&0
\eea
\bea
\label{appendix4}
I_1(B_{,tt}+B_{,zz}-2B_{,tz}-2\nu_{,t}B_{,t}-2\nu_{,z}B_{,z}-
2\nu_{,t}B_{,z}-2B_{,t}\nu_{,z})&=&0
\eea
where $I_1=9B^{4/3}e^{2\nu}+36\alpha \kappa B^{2/3} e^{2\nu}
-4\alpha(B_{,z}^2-B_{,t}^2).$

In the degenerate case where either $B_{,u}=0$ or $B_{,v}=0$ (Class I
solutions according to Taub) there are now two subcases. Either we
obtain the Class I solution of Taub \cite{taub}
which is simply flat Minkowski spacetime or we obtain,
\be
\label{jeff}
ds^2=e^{2\nu}(-4\alpha \kappa)^{-1}(-dt^2+dz^2)+(-4\alpha \kappa)
{d\chi^2\over
1-\kappa \chi^2}+\chi^2d\Omega_{II}^2)
\ee
under once again (\ref{fine}). Note once more that $\nu(t,z)$ is an
arbitrary function of $t$ and $z$ and planar symmetry is not permitted.

\end{document}